# Vibrational modes of ultrathin carbon nanomembrane mechanical resonators

*Xianghui Zhang[a],\* Reimar Waitz[b], Fan Yang[b], Carolin Lutz[b], Polina Angelova[a], Armin Gölzhäuser[a], and Elke Scheer[b]\**

[a] Fakultät für Physik, Universität Bielefeld, 33615 Bielefeld, Germany

[b] Fachbereich für Physik, Universität Konstanz, 78457 Konstanz, Germany

**\* zhang@physik.uni-bielefeld.de; elke.scheer@uni-konstanz.de**

**Abstract**

We report measurements of vibrational mode shapes of mechanical resonators made from ultrathin carbon nanomembranes (CNMs) with a thickness of approximately 1 nm. CNMs are prepared from electron irradiation induced cross-linking of aromatic self-assembled monolayers (SAMs) and the variation of membrane thickness and/or density can be achieved by varying the precursor molecule. Single- and triple-layer freestanding CNMs were made by transferring them onto Si substrates with square/rectangular orifices. The vibration of the membrane was actuated by applying a sinusoidal voltage to a piezoelectric disk on which the sample was glued. The vibrational mode shapes were visualized with an imaging Mirau interferometer using a stroboscopic light source. Several mode shapes of a square membrane can be readily identified and their dynamic behavior can be well described by linear response theory of a membrane with negligible bending rigidity. Applying Fourier transformations to the time-dependent surface profiles, the dispersion relation of the transverse membrane waves can be obtained and its linear behavior confirms the membrane model. Comparing the dispersion relation to an analytical model, the static stress of the membranes was determined and found to be caused by the fabrication process.

Nanoelectromechanical systems (NEMS) have attracted considerable attention as they have enabled the observation of the quantum regime of mechanical motion[1] and potential applications such as single-molecule or atom surface adsorption sensing[2] and signal processing.[3] Apparently, there is a growing focus on freestanding nanomembranes (NMs) that are prepared by means of a variety of emerging approaches, such as organic synthesis,[4] bottom-up self-assembly,[5,6] and top-down thinning and shaping.[7,8] Understanding of the mechanical motion of a NM in the regime of nanoscale thickness is of fundamental importance and also essential for potential applications in the NEMS as building blocks. Recently, graphene-based NEMS have been intensively studied [9-11] and their use in optomechanics[12] and signal processing[13] have also been demonstrated. $MoS_2$ mechanical resonators have been investigated experimentally and the transition from the "plate" to the "membrane" regime was identified as the number of layers decreases.[14,15] However, visualization of vibrational modes of a NM is confronted with difficulties in excitations and detections. A number of methods have evolved to visualize the vibrational mode shapes, such as scanning vibrometry,[16,17] holographic interferometry[18] and stroboscopic interferometry.[19,20] Among those, stroboscopic interferometry is capable of visualizing 3D vibrational modes with a lateral resolution in the sub-micrometer range and a vertical resolution in the sub-nanometer range, as exemplified by measurements performed on a 340 nm thick Si membrane.[20]

Carbon nanomembranes (CNMs) are a class of two-dimensional materials that are fabricated by cross-linking self-assembled monolayers (SAMs) of aromatic precursor molecules via low-energy electron irradiation.[6] Depending on the precursor molecules and the preparation conditions, the thickness and porosity of CNMs can be tailored.[21] Apart from that, the terminal functionality of CNMs is inherited from the precursor molecules but can be modified upon electron irradiation, e.g. nitro group converts to amino group,[22] which enables further molecular grafting onto amino-terminated regions.[23,24] Mechanical properties of CNMs have been characterized by bulge testing in an atomic force microscope (AFM bulge test)[25] and it was found that their Young's moduli can also be tailored in the range of 10–20 GPa, depending on the precursor molecules.[26] In contrast to graphene being a semimetal, the pristine CNMs are dielectric and structurally amorphous but can convert to well-conducting nanocrystalline graphene upon thermal annealing.[27,28] The fabrication of heterostructures of CNM and graphene suggests that CNMs can be also used as a 2D dielectric in graphene-based electronics.[29]

In this paper, we report the measurements of vibrational mode shapes of CNM-based mechanical resonators using a resonant measurement technique relying on interferometry. We show that although the reflectivity of CNMs in the visible regime is relatively low due the low thickness and small amount of material, it is sufficient to obtain interference amplitudes enabling to determine the bending wave eigenmodes and to estimate the Young's modulus.

The preparation of CNMs is schematically illustrated in Fig. 1a: (1) self-assembled monolayer (SAM) of 4´-nitro-biphenyl-4-thiol (NBPT) was prepared on a 300 nm polycrystalline Au layer thermally evaporated on mica substrates (Georg Albert PVD-Coatings, Germany); (2) the conversion of a NBPT–SAM into a NBPT–CNM (amino-terminated CNM) was achieved by exposing the SAM to low energy electrons in a high vacuum. To vary the membrane thickness and density, [1´´,4´,1´,1]-terphenyl-4-thiol (TPT) is also used for the membrane preparation. To fabricate freestanding single- or multi-layer CNMs, a transfer process has been developed, as schematically shown in Fig. 1b: (1) A poly(methyl methacrylate) (PMMA) double layer was used for spin-coating the CNM/Au/mica. (2) After separating the Au/CNM/PMMA layer from the underlying mica substrate, the sample was transferred to an etching bath where the gold substrate was dissolved. (3) The PMMA/CNM layer was directly placed onto the second CNM/Au/mica sample and the repeating of previous step yields a stack of CNMs. (4) The single-layer or multi-layer CNMs stabilized by the PMMA double layer was transferred onto a Si substrate that contains window-structured orifices (Silson Ltd., UK). (5) The dissolution of the PMMA double layer in acetone followed by the sample drying was achieved in a critical point dryer (Tousimis Autosamdri-815B). The micrograph of a freestanding triple-layer amino-terminated CNM with dimensions of $60 \times 60 \: \mu m^2$ taken by helium ion microscopy is shown in Fig. 1d.

In our experiments, the excitation of mechanical modes was conducted by piezoelectric actuation. The experimental setup is schematically illustrated in Fig. 1c. The window-structured Si substrates on which the CNM was transferred serve as a solid frame and, as we will show below, the van der Waals (vdW) interaction is strong enough to hold the CNM to the Si substrate. The sample was glued on a ring shaped piezoelectric element and such piezoelectric ring itself was glued on a massive brass block. By applying an AC voltage, thickness oscillations of the piezo were excited. The vibrations were coupled into the Si substrate and thus into the membrane. In ambient conditions, the membrane oscillations would be strongly damped by the emission of sound waves and the inertia of the air surrounding the membrane would make the analysis of the vibration more complicated.[20] Therefore the sample was located inside a vacuum cell. A pressure of about 4 mbar is achieved in our setup to suppress these complications.

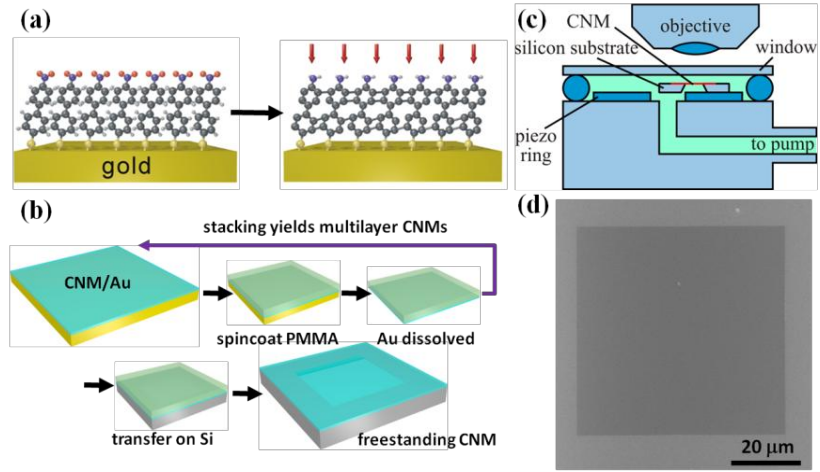

FIG. 1. (a) Schematics of the formation of NBPT–SAM on a Au substrate and the cross-linking induced by low energy electron irradiation. (b) The fabrication scheme of freestanding single- and multi-layer CNMs on a Si chip with square orifices. (c) A schematic of the experimental setup showing the mounting of the sample on a piezo ring and the measurement with an imaging Mirau interferometer. (d) The micrograph of a triple-layer amino-terminated CNM (dark grey) on a Si chip (light grey) taken by helium ion microscopy.

We use a custom made microscope objective with an integrated Mirau interferometer to measure the surface profile of the membrane. Using a stroboscopic light source synchronized with the excitation voltage, the profile was measured at 24 different phases of the oscillation. This movie of the motion of the surface profile has a time resolution of up to 80 ns, limited by the minimum pulse width of the stroboscopic light. The range of excitation frequencies from 100 Hz to 2 MHz is limited by the exposure time of the CCD camera and the bandwidth of the excitation voltage source. More details about excitation and detection of membrane waves have been described elsewhere.[20]

For a NM with a thickness of *d*, the bending stiffness can be determined from the equation $D = Ed^3/12(1-\nu^2)$ with *E* being the Young's modulus and $\nu$ the Poisson's ratio. The bending stiffness of a single-layer NBPT–CNM is estimated to be $8.5 \times 10^{-19}$ N·m, which is even four times smaller than the value $3.6 \times 10^{-18}$ N·m of graphene. Assuming that such an extremely low bending stiffness is negligible compared to the pre-tension of a CNM, the model of an infinitely thin elastic membrane without curvature contributions is used to describe the mechanics of CNM resonators. A harmonic ansatz for the out-of-plane displacement of a rectangular membrane in a Cartesian coordinate system is expressed as,

$$w(x,y,t) = \sum_{m=1}^{\infty}\sum_{n=1}^{\infty} q_{mn}(t) \sin\frac{m\pi x}{a} \sin\frac{n\pi y}{b}$$

$$x \in [0,a], y \in [0,b] \tag{1}$$

where $a$ and $b$ denote the lengths of the sides of the membrane, and $q_{mn}(t) = Q_{mn}e^{i\omega_{mn}t}$ is the temporal coefficient that describes how the out-of-plane displacement changes in time. The boundaries are considered fixed and no delamination occurs during vibration, as verified elsewhere in the case of pressure loading in the bulge test.[25,26] The spatial part $\sin\frac{m\pi x}{a}\sin\frac{n\pi y}{b}$ of this expression satisfies aforementioned boundary conditions.

According to the Rayleigh-Ritz method,[30-32] the natural frequencies of a rectangular membrane with direction dependent tensions can be expressed as

$$f_{mn} = \frac{1}{2}\sqrt{\frac{1}{\rho_s}\left(P_x \frac{m^2}{a^2} + P_y \frac{n^2}{b^2}\right)} \tag{2}$$

where $\rho_s$ is the area density, $P_x$ and $P_y$ depict the pre-tension in the $x$ and $y$ directions.

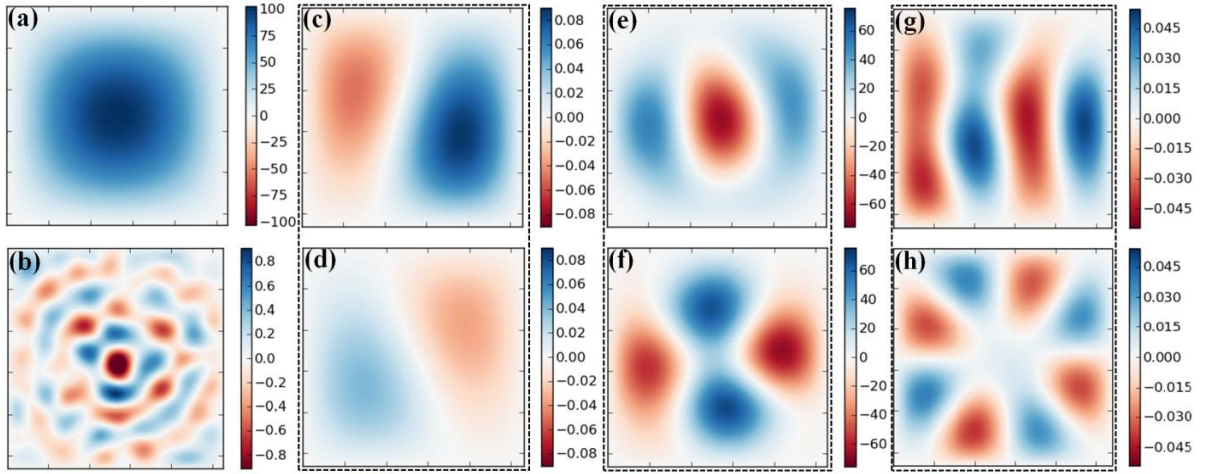

FIG. 2. Several representative vibration modes of a NBPT–CNM with dimensions of $50 \times 50\ \mu m^2$: (a) the $(1, 1)$ mode at 530 kHz; (b) a higher harmonic mode at 4.8 MHz; (c–d) two representative mode shapes of a vibrational mode at 1.2 MHz; (e–f) two mode shapes of a vibrational mode at 1.99 MHz: the former is dominated by the $(3, 1)$ mode and the latter is a superposition mode of $(3, 1)$ and $(1, 3)$ mode; (g–h) two mode shapes of a vibrational mode at 2.22 MHz: the former is dominated by the $(4, 1)$ mode and the latter is a superposition of several modes. The color scale depicts the out-of-plane displacement with a unit of nm.

From the above equations it can be seen that mode shapes predominantly depend on the pre-tension, the excitation frequencies and the lateral geometry of the CNM. Fig. 2 shows several examples of measured vibrational modes a NBPT–CNM (CNM_A, see Table I) and we label the normal modes as $(m, n)$. Fig. 2a shows the first symmetrical $(1, 1)$ mode excited at 530 kHz and its out-of-plane amplitude is about 100 nm. Fig. 2c–h show three vibrational modes and two mode shapes with a phase shift of 90º are presented for each vibrational mode. Fig. 2c shows a vibrational mode at an excitation frequency of 1.2 MHz. It is worth mentioning here that for an isotropically stressed and perfectly square membrane the orthogonal $(2, 1)$ mode and $(1, 2)$ mode are degenerate and a superposition with equal contributions from both modes leads to a diagonal nodal line. The obtained nodal line shifts slightly to the diagonal, indicating that the $(1, 2)$ mode contributes less significantly to the superposition mode than the $(2, 1)$ mode. The extremely small out-of-plane displacement of this superposition mode arises from the fact that it is a 4$^{th}$ harmonic wave generated at an excitation frequency of 300 kHz. The mode shapes in Fig. 2e and 2f were excited at 1.99 MHz, where the former appears to be dominated by the $(3, 1)$ mode and the latter a superposition mode of $(3, 1)$ and $(1, 3)$ modes. Fig. 2g and 2h show a vibrational mode that is a 6$^{th}$ harmonic wave generated at an excitation frequency of 370 kHz: the former mode shape appears to be the $(4, 1)$ mode that is slightly superimposed with other mode shapes and the latter is even more complex. The complexity of mode superposition is exemplified by the vibrational mode excited at 4.8 MHz (see Fig. 2b), where the mode shape is apparently a superposition of several orthogonal and even nonorthogonal modes.

Determination of pre-stress in a thin film has been carried out by a number of different techniques, among which resonant frequency method based on Eq. 2 shows a high degree of accuracy and reproducibility. Instead of fundamental frequencies, we could modify the Eq. 2 and obtain the pre-stress from the dispersion relation

$$f_{mn} = \frac{1}{2\pi}\sqrt{\frac{P}{\rho_s}\left[\left(\frac{m\pi}{a}\right)^2 + \left(\frac{n\pi}{b}\right)^2\right]} = \frac{1}{2\pi}\sqrt{\frac{\sigma}{\rho}} \cdot k = \sqrt{\frac{\sigma}{\rho}} \cdot \frac{1}{\lambda} \qquad (3)$$

where the wave vector is expressed as $\vec{k} = \frac{m\pi}{a}\hat{x} + \frac{n\pi}{b}\hat{y}$, and a uniform tension is assumed, i.e. $P = P_x = P_y$. Here the area density $\rho_s$ is replaced by the volumetric mass density $\rho$ and the pre-tension $P$ is replaced by the pre-stress $\sigma$.

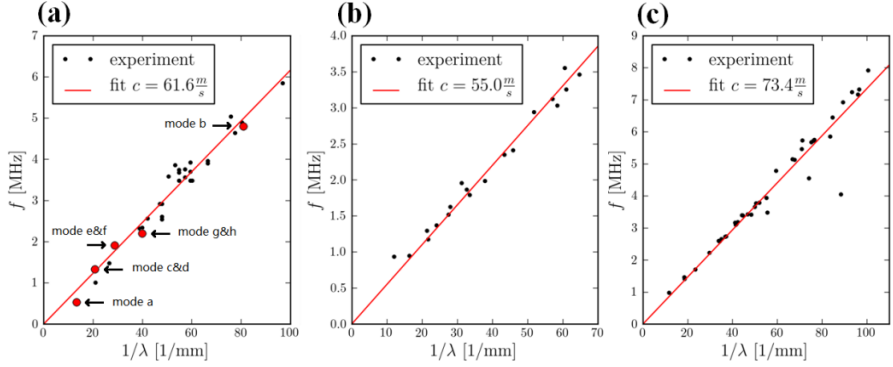

FIG. 3. (a) Dispersion relation of a single-layer NBPT–CNM resonator and the red circles indicate the five vibrational modes shown in Fig. 2. Dispersion relations of the single-layer TPT–CNM resonator and the triple-layer NBPT–CNM resonator were plotted in (b) and (c), respectively.

Fig. 3 show three dispersion relations of three CNM resonators under investigation. The data points of the dispersion relations were obtained from the measured data by searching for the highest amplitude $Q_{mn}$ for any given $\vec{k}$ vector and the so called "maximum amplitude method" is explained in detail elsewhere.[20] Each data point denotes a vibrational mode with 24 mode shapes measured at different phases. As an example, the five points labeled by red circles in Fig. 3a correspond to the five vibrational modes shown in Fig. 2. A linear relation between $f$ and $1/\lambda$ indicates that CNM resonators are indeed in the membrane regime as anticipated from the extremely low bending stiffness. The slope of a linear regression line denotes the phase velocity $v_p = f\lambda = \sqrt{\sigma/\rho}$ of the membrane wave. A phase velocity of $61.6 \pm 10.0$ m/s was obtained for a single-layer NBPT–CNM resonator. The phase velocities of $55.0 \pm 3.9$ m/s and $73.4 \pm 6.6$ m/s were calculated for a single-layer TPT–CNM resonator (CNM_B) and a triple-layer NBPT–CNM resonator (CNM_C), respectively. Most of the data points are close to the line and it is a very rare case that data points deviate significantly from the linear fitting (see the outlier in the Fig. 3c). In general, the scattering of data points in the dispersion relations can be due to uncertainties in the determination of $k$, to anisotropic stress [20,33], as well as to inhomogeneity in the area density as a result of the fabrication processes (see dot-like PMMA residue in Fig. 1d). Since the error bars for the determination of ω and $k$ are of the order of the symbol sizes, we attribute the small and systematic scatter around the linear regression line to anisotropic stress.[20,33]

TABLE I. Summary of properties of CNM resonators under investigation

| Samples | Precursor molecule | # of layers | Thickness (Å) | Size (μm²) | Density (kg/m³) | Phase velocity (m/s) | Pre-stress (MPa) |
|---|---|---|---|---|---|---|---|
| CNM_A | NBPT | 1 | 10 | 50×50 | 1301.3 | 61.6 ± 11.1 | 4.9 ± 1.8 |
| CNM_B | TPT | 1 | 12 | 78×49 | 1352.5 | 55.0 ± 3.9 | 4.1 ± 0.6 |
| CNM_C | NBPT | 3 | ~38 | 60×61 | 1027.3 | 73.4 ± 6.6 | 5.5 ± 1.0 |

To determine the stress $\sigma$ from Eq. (3), we need to estimate the density of the CNM under investigation. Assuming a pristine densely packed SAM with a structure of $(\sqrt{3} \times \sqrt{3})R30°$, a total area of ~21.6 Å² is occupied by a single molecule.[34] When the SAM is exposed to electrons, the irradiation leads to a reduction of 5–10% in the carbon density.[21] This is associated with the presence of isolated, non-linked molecules in the SAM as a consequence of steric hindrances. After the transfer process, those non-linked molecules are absent in CNMs as material got lost. Taking this into account, the membrane density was calculated and listed in Table I. Notice that the formation of voids (pores with a diameter of a few Å) in CNMs has not yet been experimentally identified, but their possible existence could be supported by the observed selectivity in gas separations.[35]

In general, the obtained pre-stresses $\sigma$ for the three CNM resonators are tensile and they are in the range of 4–6 MPa. For example, single layer TPT–CNM has a pre-stress of 4.1 MPa, which is in good agreement with the average value of 3 MPa determined from AFM bulge testing.[26] The pre-stress of the single-layer NBPT–CNM and TPT–CNM are very close to each other despite of the fact that they have different sizes as well as densities. For a triple-layer CNM, the interlayer distance is assumed to be 4 Å and this reduces the density by about 21% compared to the single-layer CNM. Though the mechanical stacking of multilayers should cause a decrease in pre-stress due to stress relaxations, the triple-layer NBPT–CNM under investigation possesses a slightly higher pre-stress and this variation is very likely due to the pre-stress of the first layer CNM. We substituted the obtained pre-tension into Eq. (2) and computed the frequencies of the first four normal modes. Due to the air inertia involved in the motion and the slightly anisotropic pre-tension in reality, a deviation between the measured and calculated $f_{mn}$ is expected. The measured $f_{11}$ is ~40% smaller than the calculated value, whereas the measured frequencies of other three normal modes are in good agreement with the values predicted by the analytical equation with a slight shift to lower frequencies by ~15%. The larger deviation of the groundmode frequency can be attributed to the fact that environmental influences are most effective at low frequencies[20]. The same extrinsic influences limit the determination of the intrinsic quality factors. From the width of the resonance curves can give the lower bound for the effective quality factor of the whole system to be $Q_{eff} > 400$.

To understand the origin of the tensile stress in a suspended CNM, the fabrication process must be analyzed step by step: (1) a compressive stress is expected when a SAM forms on the Au substrate;[36] (2) electron irradiation causes the initial compressive stress to decrease, and may even convert it into a tensile stress,[37] due to the replacement of vdW interactions among adjacent molecules by covalent bonds; (3) when a tensile-stressed CNM is coated with a PMMA layer and separated from the Au substrate, stress relaxation leads to corrugation of the PMMA layer; (4) when the PMMA/CNM is placed over a hole, the tensile strain induced by vdW interactions is very small due to the relatively large thickness (400 nm) of the PMMA layer; (5) when the PMMA layer is dissolved in acetone and the remaining CNM is dried at the critical point of $CO_2$, a readjustment of stress in the CNM is still possible.

A nearly constant tension of self-assembled nanoparticle drumhead resonators was reported by Guest et al. and vdW interactions between the sheet and the substrate are considered to account for such a constant tension.[38] The stresses caused by vdW interactions should be inversely proportional to the thickness of the membrane. Despite the fact that single- and triple-layer CNM have similar values of pre-stresses in this experiment, AFM bulge tests show that the average pre-stresses decrease with increasing the number of layers, which indicates that the vdW interaction is indeed one of the determining factors. Furthermore, we obtained different pre-stresses for CNMs prepared from different precursor molecules, in spite of their similar thicknesses. This indicates that the second fabrication step (electron irradiation) is as well important, as higher conformational degree of freedom of the precursor molecules renders more ways to cross-link the molecules and thus achieving a lower pre-stress in the CNM.[26] Therefore, we suggest that the maximum tensile stress of a CNM could be determined by the electron irradiation step, but can relax to a certain extent during the whole transfer process. However, the minimum tensile stress of a CNM is determined by vdW interactions between the CNM and the sidewall of the Si substrate. This finding indicates that the CNM-based resonators are capable of being strain engineered by modifying the fabrication process, as demonstrated in the graphene drum resonators.[39]

In summary, we have shown the fabrication of freestanding single- and triple-layer CNMs and that they can be used as mechanical resonators. Their optical reflectivity is suitable to characterize their vibrational mode shapes by optical interferometry. The extremely low mass and bending stiffness ensures that CNM resonators are in the membrane regime where the pretension of the CNM dominates over its elastic stiffness, as verified by the linear dispersion relation. The pre-stress determined from the phase velocities of the transverse waves is around 5 MPa, which is in good agreement with those obtained by AFM bulge testing. Carbon nanomembranes are fundamentally different from graphene: possessing intrinsic chemical functionality, being amorphous in structure, and being an electrically insulating material, they yet can be converted to nanocrystalline graphene; and these properties make them a novel carbon-based nanomaterial that is potentially complementary to graphene for heterogeneous devices.

The authors thank the Volkswagenstiftung, the Bundesministerium für Bildung und Forschung (BMBF), and the Deutsche Forschungsgemeinschaft through SFB767 for financial support and the SFB767 Nanomechanics Discussion Group for fruitful remarks.